\documentclass[prl,twocolumn,showpacs,preprintnumbers,amsmath,amssymb]{revtex4}

\usepackage{graphicx}
\usepackage{dcolumn}
\usepackage{bm}

\begin{document}

\title{Broadband laser cooling of trapped atoms with ultrafast pulses}

\author{B. B. Blinov}
\email{bblinov@umich.edu}
\affiliation{FOCUS Center and University of Michigan Department of Physics
Ann Arbor, MI 48109-1040}

\author{R. N. Kohn, Jr.}
\affiliation{FOCUS Center and University of Michigan Department of Physics
Ann Arbor, MI 48109-1040}

\author{M. J. Madsen}
\affiliation{FOCUS Center and University of Michigan Department of Physics
Ann Arbor, MI 48109-1040}

\author{P. Maunz}
\affiliation{FOCUS Center and University of Michigan Department of Physics
Ann Arbor, MI 48109-1040}

\author{D. L. Moehring}
\affiliation{FOCUS Center and University of Michigan Department of Physics
Ann Arbor, MI 48109-1040}

\author{C. Monroe}
\affiliation{FOCUS Center and University of Michigan Department of Physics
Ann Arbor, MI 48109-1040}

\date{\today}

\begin{abstract}
We demonstrate broadband laser cooling of atomic ions in an rf trap using ultrafast pulses
from a modelocked laser. 
The temperature of a single ion is measured by observing the size of a
time-averaged image of the ion in the known harmonic trap potential. 
While the lowest observed temperature was only about 1~K, this method efficiently cools
very hot atoms and can sufficiently localize trapped atoms to produce near diffraction-limited
atomic images.
\end{abstract}


\maketitle

Laser cooling of atoms~\cite{hansch:1975,wineland:1975}
has become a cornerstone of modern day atomic physics. 
Doppler cooling and its many extensions usually involve narrow-band, 
continuous-wave lasers that efficiently cool atoms within a narrow velocity range ($\sim1$~m/s) 
that corresponds to the radiative linewidth of a typical atomic transition. 
To increase the velocity capture range, several
laser cooling methods were investigated that modulate or effectively broaden a narrow-band 
laser~\cite{hoffnagle:1988,zhu:1991,littler:1991,ketterle:1992,calabrese:1996,atutov:1998}. 
Modelocked pulsed lasers have been used to narrow the velocity
distribution of atomic beams within several velocity classes given by the bandwidth of each 
spectral component of the frequency comb~\cite{strohmeier:1991,watanabe:1996}.
In this letter we report the demonstration of Doppler laser cooling of trapped atoms 
with individual broadband light pulses from a modelocked laser.

\begin{figure}
\includegraphics[width=1.0\columnwidth,keepaspectratio]{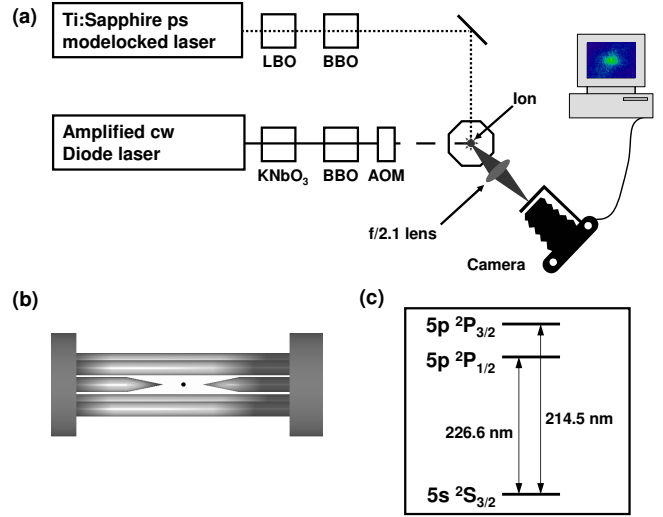}
\caption{\label{fig1}The experimental apparatus.  
(a) Frequency-quadrupled pulses from a picosecond modelocked Ti:Sapphire laser
(Spectra-Physics Tsunami) are 
tuned to the $5p~^2P_{\text{1/2}}$ transition in Cd$^+$ near 226.5~nm 
and directed onto the trapped ion.  
An amplified narrow-band diode laser is also frequency-quadrupled and tuned a few linewidths red of 
the $5P~^2P_{\text{3/2}}$ transition for initial Doppler cooling of the ion.  
An acousto-optic modulator (AOM) is used to switch on and off the narrow-band light.  
Photons emitted from the ion are collected by an f/2.1 
imaging lens and directed toward a photon-counting intensified camera.  
(b) Schematic drawing of the linear rf trap used in the experiment, with the ion position
indicated by the black dot in the middle.
(c) The relevant energy levels of Cd$^+$.}
\end{figure}

To efficiently capture and cool high-velocity atoms, it is necessary 
to achieve a laser bandwidth large enough to cover the large range of atomic Doppler shifts. 
For example, Cd$^+$ ions used in this experiment are initially created with an average kinetic energy
of order 1~eV, which corresponds to an average velocity of about 1300~m/s and 
a Doppler shift of $\Delta_{\text{D}}\sim36$~GHz.
Power broadening an atomic transition (saturation intensity $I_{\text{s}}$ and natural
linewidth $\gamma$) would require a laser intensity of
$I/I_{\text{s}}\sim\ (2\Delta_{\text{D}}/\gamma)^2$, which can be prohibitively high.
For Cd$^+$ ($\gamma/2\pi\simeq50\text{MHz}$, $I_{s}\simeq5000\text{W}/\text{m}^2$) this requires 
$I\sim10^{10}\text{W}/\text{m}^2$.
Modulating a narrow-band laser to generate high bandwidths would allow for
significantly less laser power, but it is technically difficult to generate a 100~GHz wide
modulation spectrum~\cite{zhu:1991}. On the other hand,
an ultrafast laser whose pulse is a few picoseconds long will naturally have 
a bandwidth in the above range, as well as sufficient intensity to excite the transition. 

\begin{figure*}
\includegraphics[width=2.0\columnwidth,keepaspectratio]{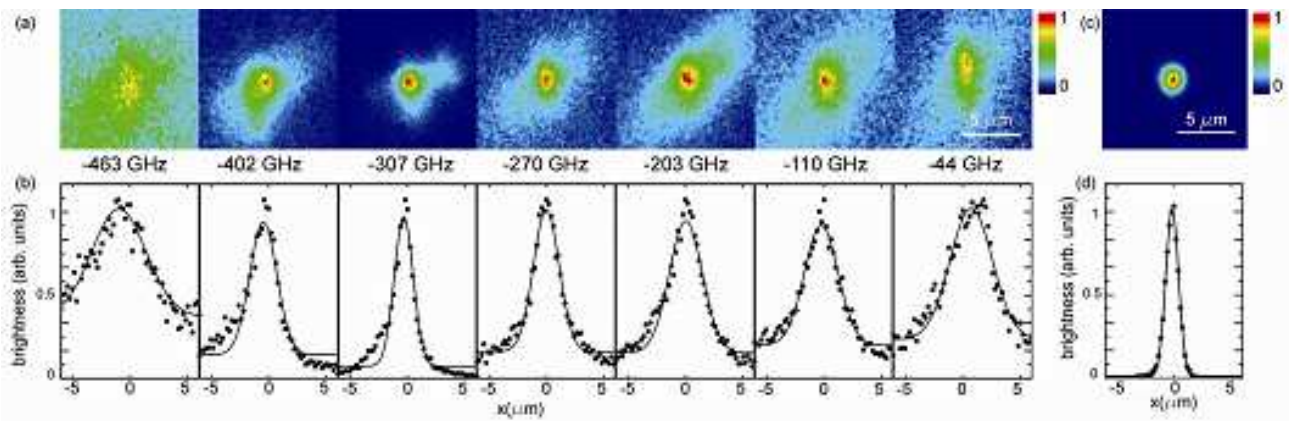}
\caption{\label{fig2}(a) Images of a single trapped ion taken at various pulsed laser
detunings $\delta/2\pi$ indicated at the bottom. The pulsed laser beam direction in each
image is diagonal from lower-left corner to upper-right corner. (b) Crossections of the
images in (a) along the vertical direction. The solid lines are Gaussian fits to the data. 
(c) An image of a narrow-band laser-cooled ion localized to $\sim$30~nm, with its crossection 
and a Gaussian fit plotted in (d).}
\end{figure*}

The laser cooling rate depends critically on the photon scatter rate, which for a pulsed laser
can be no larger than the laser repetition rate $R$
(about 80~MHz for a typical modelocked laser), given that the atom is excited with unit 
probability by each pulse. 
We assume that once excited, the atom decays back to the ground state 
faster than the time period of the modelocked pulse train $1/R$.
In this case, the atom has little memory between pulses, or equivalently, the absorption
spectrum is a single broad line of width $\Delta\sim1/\tau$ ($\tau$ is the pulse
duration) and the frequency comb of spacing $R$ has very little contrast.
 
The equilibrium temperature for broadband pulsed laser cooling of trapped atoms 
is expected to scale approximately with the laser bandwidth 
$\Delta$, and is much higher than typical narrowband laser-cooled atom temperatures. 
Still, cooling of atoms in a strong trap to these higher temperatures
can localize them to less than the diffraction limit ($\sim1\mu\text{m}$) of
typical imaging optics. This cooling may thus be sufficient for the implementation of
quantum optics applications that interface atoms with photons~\cite{simon:2003,duan:2004,duan:2005}. 
In these applications, it is necessary to mode-match single photons emitted by 
individual atoms, so the atomic image quality is important, 
while cooling to near the ground state of motion
or within the Lamb-Dicke limit is not required~\cite{leibfried:2003}.

The experimental setup is shown schematically in Fig.~\ref{fig1}.
We trap atomic cadmium ions in a linear rf (Paul) trap~\cite{moehring:2005}, 
shown in Fig.~\ref{fig1}(b). 
The spacing of four 0.5~mm diameter rods is about 1~mm, 
while the separation of the two end cap needles is about 2.6~mm. 
The strengths of the radial rf trap and the axial static trap are 
adjusted to be approximately (but not exactly) equal: 
$\omega_{\text{x}}\simeq\omega_{\text{y}}\simeq\omega_{\text{z}}\simeq 2\pi\times$0.85~MHz,
and the rf drive frequency is $\Omega_{\text{rf}}=2\pi\times35.8$~MHz.  
The trapped ions can be either Doppler-cooled with a narrow-band, cw laser tuned a few
linewidths red of the $^2S_{1/2}~-~^2\!P_{3/2}$ transition at 214.5~nm, 
or by a modelocked laser tuned red of the $^2S_{1/2}~-~^2\!P_{1/2}$ transition
at 226.5~nm. Both laser beams are oriented to have significant $k$-vector components
along each principal axis of the trap to efficiently cool all degrees of freedom of the trapped ion.
The ion fluorescence is collected by an f/2.1 lens and directed to a 
photon-counting intensified camera. The inherent chromatic aberration of the imaging system 
allows us to selectively image the 226.5~nm or the 214.5~nm fluorescence by simply adjusting the 
focus on the f/2.1 lens. 

To measure the cooling efficiency of the modelocked laser we first
Doppler-cool a single Cd$^+$ ion using the narrow-band laser, with the pulsed laser also
directed onto the ion. The narrow-band laser beam
is then turned off, and an image of the trapped ion fluorescence
is recorded using the camera, with an integration time of up to 10 minutes. 
A series of broadband laser-cooled ion images taken at various  
detunings $\delta\!=\!\omega_{\text{l}}\!-\!\omega_{\text{a}}$,
where $\omega_{\text{l}}$ is the modelocked laser central frequency,
and $\omega_{\text{a}}$ is the atomic resonance frequency, is shown in Fig.~\ref{fig2}(a).
The modelocked laser average power is held constant at 1~mW, which corresponds to individual
pulse energies of about 12.5~pJ. 
The resulting image is analyzed to measure its rms width, 
$x_{\text{im}}$, by fitting its crossection to a Gaussian distribution [Fig.\ref{fig2}(b)].

To determine the actual Gaussian rms radius $x_{\text{rms}}$ of the time-averaged ion position, 
two effects must be considered.
First is the finite resolution $x_{\text{r}}$ of the imaging optics, which we measure by 
recording an image of a narrowband laser-cooled ion [Fig\ref{fig2}(c)],
resulting in a near point-source with an estimated object size of $\sim30$~nm. Fitting 
its crossection [Fig.\ref{fig2}(d)] to a Gaussian distribution provides a good estimate of 
$x_{\text{r}}=1.15\pm0.01~\mu\text{m}$. This is about a factor of two larger than the expected 
diffraction-limited image size of about $0.55~\mu\text{m}$, which we attribute to an incomplete
correction of the spherical aberration of the f/2.1 lens.
Using properties of the convolution of Gaussian functions, we then determine the resolution-corrected
image width: $x_{\text{corr}}=\sqrt{x_{\text{im}}^2-x_{\text{r}}^2}$.

The second effect is the modulation of the ion brightness due to laser light intensity 
variation across the waist, whose measured rms width is $x_{\text{w}}~\!\!=~\!\!3.35~\!\!\pm~\!\!0.15~\!\mu\text{m}$. 
The true rms ion motion size is
$x_{\text{rms}}~=~x_{\text{w}}x_{\text{corr}}/\sqrt{x_{\text{w}}^2-x_{\text{im}}^2sin^2(\phi)}$,
where $\phi$ is the angle between the laser beam direction and the direction of ion image
crossection. We analyze the temperature in the radial and the axial directions, where
$\phi~=~\pm45^{\circ}$.

The effect of the ion micromotion (fast oscillations near the rf drive frequency) on the image 
size is negligible in our case. With the proper compensation of the background electric fields,
the micromotion amplitude is 
$x_{\text{m}}~=~(\sqrt{2}\omega/\Omega_{\text{rf}})x_{\text{rms}}\simeq0.035x_{\text{rms}}$~\cite{berkeland:1998},
where $\omega$ is the ion's secular frequency along the particular principal axis. 
Broadening of the image due to excess micromotion, which arises from an incomplete compensation of 
the background electric fields, is taken to be much smaller than the resolution $x_{\text{r}}$ of our optics.

The ion rms velocity in the trap $v_{\text{rms}}$ along a principal axis is directly proportional 
to the rms displacement: $v_{\text{rms}} = \omega x_{\text{rms}}$. 
The temperature $T$ of the ion (assuming a normal distribution of its velocity) is then given by
$k_{\text{B}}T=mv_{\text{rms}}^2$, where $k_{\text{B}}$ is the Boltzmann constant, and $m$ is the ion mass.

\begin{figure}
\includegraphics[width=1.0\columnwidth,keepaspectratio]{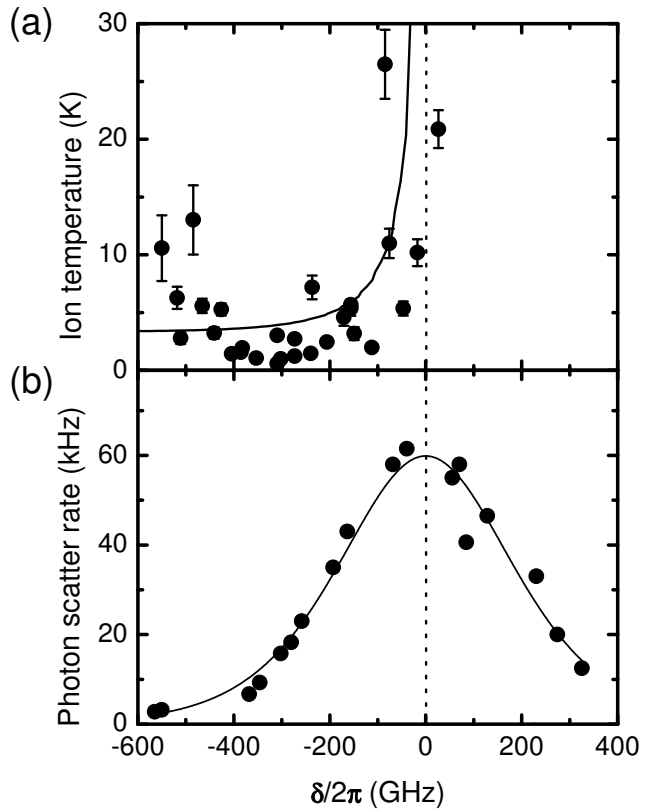}
\caption{\label{fig3}A summary of the measurements.  
(a) The measured radial ion temperature is plotted against the pulsed laser detuning $\delta$. The
solid line represents the theoretically predicted temperature [Eq.~\ref{temp}].
(b) Photon scatter rate from a single, cold ion is plotted against the pulsed laser detuning. 
The vertical dashed line indicates the atomic resonance position, corresponding to the
wavelength 226.57~nm. The solid line
is a fit to the data using $sech^2$ spectrum [Eq.~\ref{abs}], indicating 
$\Delta\sim420$~GHz and $\tau\simeq1.3$~ps}
\end{figure}

The summary of our results is shown in Fig.~\ref{fig3}. 
For the ion temperature data in Fig.~\ref{fig3}(a), each point is measured using the 
procedure described above~\cite{crossection:footnote}. 
The absorption lineshape in Fig.~\ref{fig3}(b) is taken in a separate experiment 
by measuring the fluorescence rate of a single cold ion under a pulsed laser average power of 1~mW. 
For this, a 100~$\mu s$ narrowband laser-cooling cycle is interlaced with a 200~$\mu s$ period when 
only the pulsed laser light is incident on the ion and the ion fluorescence is collected. 
There is a wide range of pulsed laser detunings in Fig.~\ref{fig3}(a) 
for which the ion temperature is well below
5~K, reaching as low as 1~K. These detunings correspond to the region of high slope
in the absorption line curve, as expected in Doppler cooling~\cite{metcalf}.
The ion temperature increases sharply as $\delta$ approaches zero; 
it also grows significantly on the
far-red side of the resonance, where the cooling rate is very slow due to low photon scatter rate, 
while additional background heating~\cite{blumel:1989,turchette:2000} 
presumably increases the equilibrium temperature of the ion.  

The bandwidth of the laser pulses used in the experiment
is measured to be $\Delta\sim2\pi\times420$~GHz, as shown in Fig~\ref{fig3}(b),
which is almost three orders of magnitude larger than 
the linewidth $\gamma/2\pi\simeq50.5$~MHz of the $5p^2P_{\text{1/2}}$ Cd$^+$
excited state~\cite{moehring:2005}. Thus, the velocity-dependent 
(frictional) force that leads to cooling arises from the laser line shape
rather than the atomic line shape. 

The cooling mechanism can still be understood in terms similar to conventional Doppler 
cooling~\cite{metcalf}. 
The probability of absorbing a photon by the ion is velocity-dependent, due to Doppler shifts. 
With the laser central frequency tuned to the red of the atomic resonance 
($\delta<0$), 
the atom has higher probability of absorbing a photon when it is moving towards 
the laser beam, experiencing a blue Doppler shift. This absorption reduces the atom 
velocity in the direction of motion. The following spontaneous emission is random and equally 
likely in any direction; thus, the net effect of absorption and emission is to lower 
the kinetic energy of the atom. 
For a bound atom, as in the case of an ion in an rf trap, only one
cooling laser beam is necessary, provided that its $k$-vector has components along all
three trap principal axes~\cite{wineland:1979,itano:1982}. The expressions derived for cooling rate 
and the cooling limit remain the same for a free atom and three pairs of counter-propagating
cooling laser beams.

The average force due to scattering of photons from the laser beam 
experienced by the atom along a principal axis in the trap in this configuration is:
\begin{equation}
\label{force}
F = \Delta pRP_{\text{exc}},
\end{equation}
where $\Delta p = \hbar k/\sqrt{3}$ is the average momentum 
kick along the principal axis from each 
photon absorption, with $k$ being the photon's wavenumber, 
$R$ the modelocked laser repetition rate, and we assume that $\vec{k}$ has
equal components along each trap axis~\cite{itano:1982}.
The atomic excitation probability $P_{\text{exc}}$ can be derived analytically
for hyperbolic secant pulses $E_0sech(\pi t/\tau)$~\cite{rosen:1932} 
of electric field amplitude $E_0$ and duration $\tau$, expected from the modelocked laser:
\begin{equation}
\label{abs}
P_{\text{exc}} = sin^2(\theta/2)sech^2(\tau(\delta+kv)/2),
\end{equation}
where $\theta$ is the Rabi rotation angle from a resonant laser pulse,
$\tau$ is the pulse duration, and $v$ is the atom velocity component along the
laser beam.

For small values of $v$, the force [Eq.~\ref{force}] becomes
\begin{equation}
F \simeq F_0+\beta v,
\end{equation}
where the offset force $F_0=\Delta pRsin^2(\theta/2)sech^2(\tau\delta/2)$ shifts 
the equilibrium position of the trapped atom by
$F_0/(m\omega^2)\sim1$~nm in our trap~\cite{itano:1982}, and 
$\beta v=\Delta pk\tau Rsin^2(\theta/2)sech^2(\tau\delta/2)tanh(\tau\delta/2)v$
is a damping force for $\delta<0$, corresponding to red detuning 
of the laser, with the cooling rate $\beta/m$. In our experiment, the maximum
cooling rate $\beta/m\simeq2~\text{sec}^{-1}$.

This cooling is opposed by diffusion heating resulting from photons emitted by the atom 
in random directions:
\begin{equation}
\label{heat}
D = \frac{1}{3}(2E_{\text{r}})RP_{\text{exc}},
\end{equation}
where $E_{\text{r}}=\frac{(\hbar k)^2}{2m}$ is the photon recoil energy,
and the factor of $1/3$ is due to the diffusion energy equally distributed between the three
degrees of freedom~\cite{itano:1982}.
Equating the cooling power $\beta~\!\!v^2$ to the heating power $D$, we can
find the equilibrium temperature of the atom:
\begin{equation}
\label{temp}
T = \frac{\hbar}{\sqrt{3}\tau k_{\text{B}}}\frac{1}{tanh(\tau\delta/2)},
\end{equation}
where substitutions for $\beta$ and $P_{\text{exc}}$ have been made.

The predicted ion temperature $T$ corresponding to Eq.~\ref{temp} is plotted in Fig.~\ref{fig3}(a) 
in a solid line. Note that this line is not a fit to the data;
rather, it is a theoretical prediction based on the laser and
trap parameters used in the experiment. The theory and experiment are in a good 
agreement for the radial measurements, while the measured axial temperatures (not shown in Fig.~\ref{fig3}) were consistently lower
than the theory~\cite{crossection:footnote}.

It is important to point out that the lifetime of the Cd$^+$ $5p~^2P_{1/2}$ excited state is only 
3.146~ns~\cite{moehring:2005}, 
while the period of the laser pulses is 12.5~ns. Thus, by the time the next laser pulse arrives,
the excited state population is only about 2\%. This cooling process is then
primarily due to absorbing single photons from individual pulses, and not due to an optical frequency
comb effect~\cite{strohmeier:1991,watanabe:1996,kielpinski:2003}. 
For optimal cooling of a given atomic species,
the pulsed laser repetition rate should be of the order of the atom's excited state linewidth,
while the energy in each laser pulse should correspond to $P_{\text{exc}}\simeq1$ 

In summary, we have observed and quantified laser-cooling of a single, 
trapped atom by broadband, modelocked laser pulses.
The cooling is efficient, while the lowest temperatures are in single digits Kelvin.
Lower temperatures should be possible if longer modelocked laser pulses are used,
as predicted by Eq.~\ref{temp}, where the final atom temperature scales approximately as the 
inverse of the pulse duration $\tau$. Such cooling of ions in strong rf traps 
localizes them to under 1~$\mu\text{m}$, which allows diffraction-limited ion imaging.

This work was supported by the U.S. National Security Agency and Advanced Research 
and Development Activity under Army Research Office contract DAAD19-01-1-0667 and 
the National Science Foundation Information Technology Research Program.


\begin{thebibliography}{24}
\bibitem[{cro()}]{crossection:footnote}
\bibinfo{note}{Here, we present the analysed temperature data for radial
  direction in the trap only (vertical in Fig.2(a)). In the axial direction
  (horizontal in Fig.2(a)), the observed temperature is about 5 times lower.
  This suggests that our simple theoretical model does not fully describe the
  cooling mechanism.}
\end{thebibliography}
\end{document}